\documentclass[aps,preprint,amsmath,amssymb,t1,11pt]{revtex4}
\usepackage{graphicx}
\usepackage{epstopdf}
\usepackage{xcolor}
\usepackage{slashed}
\begin{document}
\title{Sensitivity of anomalous quartic gauge couplings via tri-photon production at FCC-hh}
\author{A. Senol}
\email[]{senol_a@ibu.edu.tr} 
\author{H. Denizli}
\email[]{denizli_h@ibu.edu.tr}
\author{C. Helveci}
\email[]{}
\affiliation{Department of Physics, Bolu Abant Izzet Baysal University, 14280, Bolu, T\"{u}rkiye.}

\begin{abstract}
A direct investigation of the self-couplings of gauge bosons, completely described by the non-Abelian gauge symmetry of the Standard Model, is extremely valuable in understanding the gauge structure of the SM. Any deviation from the SM predictions on gauge boson self-coupling is to give a hint at the existence of a new physics beyond the SM, which is defined with a modification of the self-interactions using an effective field theory approach. In this paper, we present a detailed Monte Carlo study searching for anomalous quartic gauge dimension-8 couplings related to $\gamma\gamma\gamma\gamma$ and $\gamma\gamma\gamma Z$ vertices at the future hadron-hadron collider (FCC-hh) via tri-photon production at a 100 TeV center of mass energy with an integrated luminosity L$_{int}$=30 ab$^{-1}$. Events that have been parton showered and include detector effects are analyzed with a Toolkit for Multivariate Data Analysis (TMVA) using a boosted decision tree to help distinguish between signal and background events to achieve the best sensitivities on anomalous quartic gauge couplings. Our obtained results reveal that the limits on anomalous quartic gauge couplings $f_{T8}/\Lambda^{4}$ and $f_{T9}/\Lambda^{4}$ at 95\% C.L. without systematic errors are about three orders of magnitude stronger compared to the best current experimental limits reported by the ATLAS collaboration at the LHC. Considering a realistic systematic uncertainty such as 10\% from possible experimental sources, our obtained limits of anomalous quartic couplings get worse by about one order of magnitude compared to those without systematic uncertainty but are still two orders of magnitude better than those recently reported by ATLAS.
\end{abstract}
%\pacs{30.15.Ba}
\maketitle

\section{Introduction}
The self-interactions of gauge bosons are defined by the non-Abelian structure of the electroweak (EW) sector within the framework of the SM. Therefore, precise measurements of their couplings play a key role in exploring new physics beyond the Standard Model (SM) as well as shedding light on the nature of the EW symmetry-breaking mechanism. Since there have been no deviations observed in the SM prediction of gauge boson self-couplings from the LHC data available so far, these couplings are used to search for indirect signs of new physics with the Effective Field Theory (EFT) approach as a model-independent tool. In this approach, new particles cannot be directly produced, and new physics emerges only as new interactions between known SM particles supplemented by operators higher than dimension-4 under the condition that the energy scale of the new physics is much higher than the center of mass energy \cite{Buchmuller:1985jz,Hagiwara:1993ck}. Assuming the Triple Gauge boson Couplings (TGC) and Higgs couplings are constrained by other processes with better experimental accuracy, such as di-boson and Higgs on-shell measurements, the EFT of new physics which modifies interactions between electroweak gauge bosons, includes 18 different dimension-8 operators \cite{Eboli:2003nq, Eboli:2006wa,Ellis:2019zex,Ellis:2020ljj,Ellis:2022zdw,Ellis:2023ucy}.

The current experimental information on dimension-8 operators for anomalous Quartic Gauge Couplings (aQGC) arises from three production channels: the production of triple gauge bosons \cite{CMS:2014cdf, ATLAS:2015ify, CMS:2017tzy, ATLAS:2017bon, CMS:2019mpq, CMS:2021jji}
, inclusive Vector Boson Scattering (VBS) di-bosons \cite{ATLAS:2014jzl, ATLAS:2016nmw, ATLAS:2016snd, CMS:2017zmo, CMS:2017fhs, CMS:2019qfk, CMS:2020ioi, CMS:2020gfh, CMS:2020fqz, CMS:2020ypo, CMS:2021gme, ATLAS:2022nru}, and exclusive di-bosons \cite{CMS:2013hdf, ATLAS:2016lse} by the ATLAS and CMS collaborations at the LHC. The triple gauge bosons production opens up multiple potential decay channels categorized by the number of charged leptons in the final state. On the other hand, the VBS di-boson production topology is characterized by two forward jets and two vector bosons. In the exclusive di-boson production, those two very forward jets are not detected. Since tri-boson processes are normally suppressed with 1/s in the s-channel propagator, the sensitivity of tri-boson production processes to aQGCs is not better than VBS processes, as observed in the current experimental results obtained at the LHC. There have been many phenomenological studies on aQGC via different tri-boson production mechanisms at hadron colliders \cite{Yang:2012vv,Ye:2013psa,Degrande:2013yda,Gutierrez-Rodriguez:2013eya,ATLAS:2013uod,Wen:2014mha, Kurova:2017int, Jager:2017owh, Oh:2019wyq, Zhu:2020ous, Senol:2021wza, Guo:2021jdn,Gutierrez-Rodriguez:2021hip}. Tri-photon production among the tri-boson production mechanisms in the hadron colliders provides an ideal platform to search for deviations from SM since it is rare in the SM and involves only pure electroweak interaction contributions at tree level \cite{deCampos:1998xx, Gonzalez-Garcia:1998zdk, ATLAS:2017lpx, Denizli:2019ijf, Denizli:2020wvn}. Furthermore, quartic neutral gauge vertices ($\gamma\gamma\gamma\gamma$ and $\gamma\gamma\gamma Z$) that contribute $pp\to\gamma\gamma\gamma$ to contain at least three photons allow electromagnetic gauge invariance alone only when they arise from dimension-8 (or higher) operators.

Given that the precision of the aQGCs scales with the center of mass energy of the colliders, multi-TeV energy colliders become an ideal platform for studying the self-interactions of gauge bosons. One of the multi-TeV collider projects currently under consideration by CERN is the Future Circular Collider (FCC) facility, which would be planned to build in a 100 km tunnel and designed to deliver $pp$, $e^+e^-$ and ep collisions \cite{FCC:2018byv}. After completing the missions of the LHC and High-Luminosity LHC (HL-LHC) physics programmes, the FCC facility, which has the potential to search for a wide parameter range of new physics, will come to the fore as an energy frontier collider project. One of the FCC options, the FCC-hh, is designed to provide proton-proton collisions at the proposed 100 TeV centre-of-mass energy with an integrated luminosity of 30 ab$^{-1}$ \cite{FCC:2018vvp}.

In this study, we investigate the sensitivity of dimension-8 effective operators related to the anomalous quartic $\gamma\gamma\gamma\gamma$ and $\gamma\gamma\gamma Z$ vertices through the tri-photon process at the FCC-hh collider. Realistic detector effects are included in the production of the signal and background processes. The Toolkit for Multivariate Data Analysis (TMVA) and its implementation of the boosted decision trees (BDTs) method, one of the widely used and rapidly developed machine learning methods in particle physics, is used to improve the separation between signal and relevant backgrounds. After obtaining the sensitivities on aQGC couplings, we compare them with the current experimental limits and comment on the numerical results.
\section{Effective dimension-8 interactions for aQGC}

aQGCs can be constructed by either linear or non-linear representations using an EFT framework. In the linear representation, the electroweak symmetry can be broken by the conventional SM Higgs mechanism while in the non-linear representation, the Higgs couplings are treated as additional free parameters. Since no deviations from the light SM Higgs predictions have been observed so far, it is more preferable to study an effective Lagrangian in a model-independent way for the anomalous quartic couplings, assuming that new physics beyond the SM still keeps $SU(2)_L\times U(1)$ gauge invariance in linear representation \cite{Eboli:2006wa,Degrande:2013rea,Perez:2018kav,Almeida:2020ylr}. In order to construct a linearly parameterized effective Lagrangian for aQGC in the EFT framework, we start with the covariant derivative acting on the Higgs doublet field;
\begin{equation}
  D_\mu  \begin{pmatrix}
    0 \\ \frac{v+ H}{\sqrt{2}} 
  \end{pmatrix} \equiv 
\left( \partial_\mu + i \frac{g'}{2} B_\mu  + i g W_\mu^i \frac{\tau^i}{2} \right)  \begin{pmatrix}
    0 \\ \frac{v+ H}{\sqrt{2}} 
  \end{pmatrix} 
  \label{eq:def_dmu}
\end{equation}
This covariant derivative is normalized such that
\begin{equation*}
[D_\mu,D_\nu]=\widehat{W}_{\mu\nu}+\widehat{B}_{\mu \nu}
\end{equation*}
where $\widehat{W}_{\mu\nu} =  W_{\mu\nu}^j\frac{\sigma^j}{2}$ and $ \widehat{B}_{\mu \nu} = B_{\mu\nu}$  are field strength tensors of the $SU(2)$ and $U(1)$ with following definition
\begin{eqnarray}
W_{\mu\nu}&=&\frac{i}{2} g \tau^i ( \partial_{\mu} W_{\nu}^i-\partial_{\nu}W_{\mu}^i+g \epsilon_{ijk} W_{\mu}^j W_{\nu}^k)\nonumber\\
B_{\mu\nu}&=&\frac{i}{2}g'(\partial_{\mu}B_{\nu}-\partial_{\nu}B_{\mu}).
\end{eqnarray}
and $\tau^i (i=1,2,3)$ represent Pauli matrices.  

The dimension-8 operators are separated into three classes: i) covariant derivatives of Higgs doublet only ($\mathcal O_{S,j}$), ii) two field strength tensors and two derivatives of Higgs doublet ($\mathcal O_{M,j}$) and iii) field strength tensors only  ($\mathcal O_{T,j}$). Therefore, the Lagrangian of the dimension-8 operators contributing to aQGCs can be written as
\begin{eqnarray}\label{lag}
\mathcal{L}_{eff}=\mathcal{L}_{SM}+\sum_{j=0}^{1}\frac{f_{S,j}}{\Lambda^4}\mathcal{O}_{S,j}+\sum_{j=0}^{7}\frac{f_{M,j}}{\Lambda^4}\mathcal{O}_{M,j}+\sum_{j=0\atop \ j\neq 3}^{9}\frac{f_{T,j}}{\Lambda^4}\mathcal{O}_{T,j}
\end{eqnarray}
where $\Lambda$ is the scale of new physics, and $f_{S,j}$, $f_{M,j}$ and  $f_{T,j}$ represent coefficients of relevant effective operators. These coefficients are zero in the SM prediction. The explicit forms of 18 dimension-8 quartic gauge operators are given in Refs. \cite{Eboli:2006wa, Degrande:2013rea,Perez:2018kav,Almeida:2020ylr}:
\[
\begin{array}{ll}
\mathcal{O}_{S0}=[(D_{\mu}\Phi)^{\dag}D_{\nu}\Phi]\times[(D^{\mu}\Phi)^{\dag}D^{\nu}\Phi],&\mathcal{O}_{T0}=\textrm{Tr}[\widehat{W}_{\mu\nu}\widehat{W}^{\mu\nu}]\times \textrm{Tr}[\widehat{W}_{\alpha\beta}\widehat{W}^{\alpha\beta}]\\
\mathcal{O}_{S1}=[(D_{\mu}\Phi)^{\dag}D^{\mu}\Phi]\times[(D_{\nu}\Phi)^{\dag}D^{\nu}\Phi],&\mathcal{O}_{T1}=\textrm{Tr}[\widehat{W}_{\alpha\nu}\widehat{W}^{\mu\beta}]\times \textrm{Tr}[\widehat{W}_{\mu\beta}\widehat{W}^{\alpha\nu}]\\
\mathcal{O}_{S2}=[(D_{\mu}\Phi)^{\dag}D_{\nu}\Phi]\times[(D^{\nu}\Phi)^{\dag}D^{\mu}\Phi],&\mathcal{O}_{T2}=\textrm{Tr}[\widehat{W}_{\alpha\mu}\widehat{W}^{\mu\beta}]\times \textrm{Tr}[\widehat{W}_{\beta\nu}\widehat{W}^{\nu\alpha}]\\
\mathcal{O}_{M0}=\textrm{Tr}[\widehat{W}_{\mu\nu}\widehat{W}^{\mu\nu}]\times[(D_{\beta}\Phi)^{\dagger}D^{\beta}\Phi],&\mathcal{O}_{T5}=\textrm{Tr}[\widehat{W}_{\mu\nu}\widehat{W}^{\mu\nu}]\times \widehat{B}_{\alpha\beta}B^{\alpha\beta}\\
\mathcal{O}_{M1}=\textrm{Tr}[W_{\mu\nu}\widehat{W}^{\nu\beta}]\times[(D_{\beta}\Phi)^{\dagger}D^{\mu}\Phi],&\mathcal{O}_{T6}=\textrm{Tr}[\widehat{W}_{\alpha\nu}\widehat{W}^{\mu\beta}]\times \widehat{B}_{\mu\beta}\widehat{B}^{\alpha\nu}\\
\mathcal{O}_{M2}=[\widehat{B}_{\mu\nu}\widehat{B}^{\mu\nu}]\times[(D_{\beta}\Phi)^{\dagger}D^{\beta}\Phi],&\mathcal{O}_{T7}=\textrm{Tr}[\widehat{W}_{\alpha\mu}\widehat{W}^{\mu\beta}]\times \widehat{B}_{\beta\nu}\widehat{B}^{\nu\alpha}\\
\mathcal{O}_{M3}=[\widehat{B}_{\mu\nu}\widehat{B}^{\nu\beta}]\times[(D_{\beta}\Phi)^{\dagger}D^{\mu}\Phi],&\mathcal{O}_{T8}=[\widehat{B}_{\mu\nu}\widehat{B}^{\mu\nu}\widehat{B}_{\alpha\beta}\widehat{B}^{\alpha\beta}]\\
\mathcal{O}_{M4}=[(D_{\mu}\Phi)^{\dagger}\widehat{W}_{\beta\nu} D^{\mu}\Phi]\times \widehat{B}^{\beta\nu},&\mathcal{O}_{T9}=[\widehat{B}_{\alpha\mu}\widehat{B}^{\mu\beta}\widehat{B}_{\beta\nu}\widehat{B}^{\nu\alpha}] \\
\mathcal{O}_{M5}=[(D_{\mu}\Phi)^{\dagger}\widehat{W}_{\beta\nu} D^{\nu}\Phi]\times \widehat{B}^{\beta\mu}+h.c.,&\\
%\mathcal{O}_{M6}=[(D_{\mu}\Phi)^{\dagger}\widehat{W}_{\beta\nu}\widehat{W}^{\beta\nu} D^{\mu}\Phi],&\\
\mathcal{O}_{M7}=[(D_{\mu}\Phi)^{\dagger}\widehat{W}_{\beta\nu}\widehat{W}^{\beta\mu} D^{\nu}\Phi].&
\end{array}
\]
A complete list of corresponding quartic gauge boson vertices modified by dimension-8 operators is given in Table \ref{tab5}.
\begin{table}
\caption{Quartic gauge boson vertices modified by the related dimension-8 operator are marked with $\checkmark$}
\label{tab5}
\begin{ruledtabular}
\begin{tabular}{cccccccccc}
&$WWWW$ &$WWZZ$ &$ZZZZ$  &$WW\gamma Z$ &$WW\gamma\gamma$ & $ZZZ\gamma$ & $ZZ\gamma\gamma$ & $Z\gamma\gamma\gamma$ & $\gamma\gamma\gamma\gamma$ \\ \hline
$\mathcal{O}_{S0},\mathcal{O}_{S1}$ & $\checkmark$& $\checkmark$& $\checkmark$& $$& $$& $$& $$& $$& $$\\
$\mathcal{O}_{M0},\mathcal{O}_{M1},\mathcal{O}_{M6}\mathcal{O}_{M7}$ & $\checkmark$& $\checkmark$& $\checkmark$& $\checkmark$& $\checkmark$& $\checkmark$& $\checkmark$& $$& $$\\
$\mathcal{O}_{M2},\mathcal{O}_{M3},\mathcal{O}_{M4},\mathcal{O}_{M5}$ & $$& $\checkmark$& $\checkmark$& $\checkmark$& $\checkmark$& $\checkmark$& $\checkmark$& $$& $$\\
$\mathcal{O}_{T0},\mathcal{O}_{T1},\mathcal{O}_{T2}$ & $\checkmark$& $\checkmark$& $\checkmark$& $\checkmark$& $\checkmark$& $\checkmark$& $\checkmark$& $\checkmark$& $\checkmark$\\
$\mathcal{O}_{T5},\mathcal{O}_{T6},\mathcal{O}_{T7}$ & $$& $\checkmark$& $\checkmark$& $\checkmark$& $\checkmark$& $\checkmark$& $\checkmark$& $\checkmark$& $\checkmark$\\
$\mathcal{O}_{T8},\mathcal{O}_{T9}$ & $$& $$& $\checkmark$& $$& $$& $\checkmark$& $\checkmark$& $\checkmark$& $\checkmark$\\
\end{tabular}
\end{ruledtabular}
\end{table}
Due to phase space characterisation of our interested signal process $pp\to\gamma\gamma\gamma$, the best sensitivity for anomalous quartic couplings of the $\gamma\gamma\gamma Z$ and $\gamma\gamma\gamma\gamma$ vertices is attainable. Since only field strength tensor type operators ($\mathcal{O}_{T,x}$) receive contributions to the $\gamma\gamma\gamma Z$ and $\gamma\gamma\gamma\gamma$ quartic vertices, we consider $\mathcal{O}_{T,8}$ and $\mathcal{O}_{T,9}$ couplings to represent all types of $\mathcal{O}_{T,x}$ operators.

The explicit expression of the anomalous quartic $\gamma\gamma\gamma Z$ and $\gamma\gamma\gamma\gamma$ vertices as a function of $f_{T8}/\Lambda^{4}$ and $f_{T9}/\Lambda^{4}$ couplings as follows 
\begin{eqnarray}
 \Gamma_{\gamma\gamma\gamma Z}^{\mu\nu\alpha\beta}(p_1,p_2,p_3,p_4)&=&-8s_W c_W^3 [4\frac{f_{T8}}{\Lambda^{4}}G(f_{T8})+\frac{f_{T9}}{\Lambda^{4}}G(f_{T9})]  \, \\
  \Gamma_{\gamma\gamma\gamma \gamma}^{\mu\nu\alpha\beta}(p_1,p_2,p_3,p_4)&=&8c_W^4 [4\frac{f_{T8}}{\Lambda^{4}}G(f_{T8})+\frac{f_{T9}}{\Lambda^{4}}G(f_{T9})]
\end{eqnarray} where sine (cosine) of the weak mixing angle ($\theta_W$) denotes by $s_W(c_W)$,  $G(f_{T8})$ and $G(f_{T9})$ are defined as follows;
\begin{eqnarray*}
G(f_{T8})&=&(p_1\cdot p_2)\Big[g^{\mu\nu}\big(g^{\alpha\beta}(p_3\cdot p_4)-p_3^{\beta}p_4^{\alpha}\big)\Big]-(p_3\cdot p_4)g^{\alpha\beta}p_1^{\nu}p_3^{\mu}+p_1^{\nu}p_2^{\mu}p_3^{\beta}p_4^{\alpha}+p_1^{\beta}p_2^{\alpha}p_3^{\nu}p_4^{\mu}\\&+&p_1^{\alpha}p_2^{\beta}p_3^{\mu}p_4^{\nu}+\color{red}\Big[\color{black}(p_1\cdot p_3)g^{\mu\alpha}\big(g^{\nu\beta}(p_2\cdot p_4)-p_2^{\beta}p_4^{\alpha}\big)-(p_2\cdot p_3)g^{\nu\alpha}p_1^{\beta}p_4^{\mu}\color{red}\Big]\color{black}+\color{red}\Big[\color{black}
 p_3\leftrightarrow p_4, \alpha \leftrightarrow \beta
\color{red}\Big]\color{black}
\end{eqnarray*}
\begin{eqnarray*}
 G(f_{T9})&=&(p_1\cdot p_2)\Big[g^{\mu\alpha}\big(g^{\nu\beta}(p_3\cdot p_4)-p_3^{\beta}p_4^{\nu}\big)+g^{\mu\beta}\big(g^{\nu\alpha}(p_3\cdot p_4)-p_3^{\nu}p_4^{\alpha}\big)+g^{\alpha\beta}(p_3^{\nu}p_4^{\mu}+p_3^{\mu}p_4^{\nu})\\&-&g^{\nu\beta}p_3^{\mu}p_4^{\alpha}-g^{\nu\alpha}p_3^{\beta}p_4^{\mu}\Big]+(p_3\cdot p_4)\Big[g^{\mu\nu}(p_1^{\beta}p_2^{\alpha}+p_1^{\alpha}p_2^{\beta})-g^{\mu\beta}p_1^{\nu}p_2^{\alpha}-g^{\mu\alpha}p_1^{\nu}p_2^{\beta}-g^{\nu\beta}p_1^{\alpha}p_2^{\mu}\\&-&g^{\nu\alpha}p_1^{\beta}p_2^{\mu}\Big]
  +p_1^{\nu}p_2^{\beta}p_3^{\mu}p_4^{\alpha}+p_1^{\nu}p_2^{\alpha}p_3^{\beta}p_4^{\mu}+p_1^{\beta}p_2^{\mu}p_3^{\nu}p_4^{\alpha}+p_1^{\beta}p_2^{\alpha}p_3^{\mu}p_4^{\nu}+p_1^{\alpha}p_2^{\beta}p_3^{\nu}p_4^{\mu}+p_1^{\alpha}p_2^{\mu}p_3^{\beta}p_4^{\nu}
 \\&+&\color{red}\bigg[\color{black}(p_1\cdot p_3)
\Big[g^{\alpha\beta}\big(g^{\mu\nu}(p_2\cdot p_4)-p_2^{\mu}p_4^{\nu}\big)+g^{\mu\beta}\big(g^{\nu\alpha}(p_2\cdot p_4)-p_2^{\alpha}p_4^{\nu}\big)+g^{\nu\beta}(p_2^{\mu}p_4^{\alpha}+p_2^{\alpha}p_4^{\mu})\\&-&g^{\mu\nu}p_2^{\beta}p_4^{\alpha}-g^{\nu\alpha}p_2^{\beta}p_4^{\mu}\Big]
+(p_2\cdot p_3)
\Big[ g^{\mu\beta}\big(p_1^{\nu}p_4^{\alpha}+p_1^{\alpha}p_4^{\nu}\big)-g^{\mu\nu}p_1^{\beta}p_4^{\alpha}-g^{\nu\beta}p_1^{\alpha}p_4^{\mu}-g^{\alpha\beta}p_1^{\nu}p_4^{\mu}\\&-&g^{\mu\alpha}p_1^{\beta}p_4^{\nu}\Big]\color{red}\bigg]\color{black}+
\color{red}\bigg[\color{black} p_3\leftrightarrow p_4, \alpha \leftrightarrow \beta \color{red}\bigg]
\end{eqnarray*}

\begin{figure}
\includegraphics[scale=0.5]{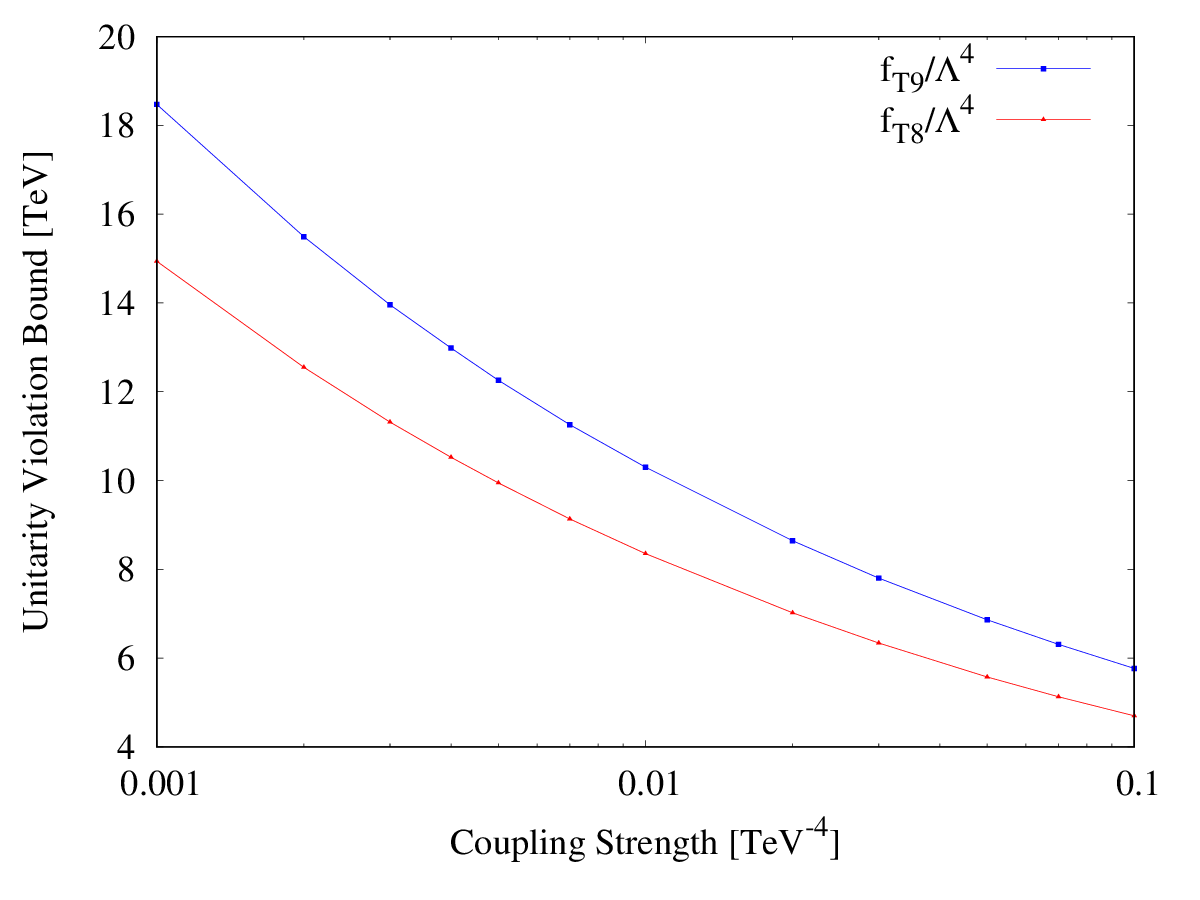} 
\caption{The UV bound as a function of aQGC $f_{T8}/\Lambda^{4}$ and $f_{T9}/\Lambda^{4}$ via on-shell VV ($V = W / Z / \gamma$) scattering processes.\label{fig1}}
\end{figure}

Since the amplitude predicted by higher-dimensional operators will eventually violate unitarity with the increase at the center of mass energy (called the unitarity bound), it becomes even more important to avoid non-physical contributions from unitarity-violated regions. One of the methods for avoiding unitarity bound for the dimension-8 operators is to implement a dipole form factor, ensuring unitarity at high energies such as: 
\begin{eqnarray}
FF=\frac{1}{(1+\hat s/\Lambda_{FF}^2)^2}
\end{eqnarray}
where $\hat s$ is the partonic center-of-mass energy, $\Lambda_{FF}$ represents the form factor cut-off scale. The $\Lambda_{FF}$ can be calculated with the form factor tool VBFNLO 2.7.1 \cite{Arnold:2008rz} for a given input of anomalous quartic gauge boson coupling parameters. The VBFNLO utility determines the form factor using the amplitudes of on-shell VV scattering processes and computes the zeroth partial wave of the amplitude. The real part of the zeroth partial wave must be below 0.5, which is called the unitarity criterion. All channels have the same electrical charge $Q$ in $VV\to VV$ scattering ($V = W / Z / \gamma$)  are combined in addition to individual checks on each channel of the $VV$ system.  The calculated Unitarity Violation (UV) bounds using the form factor tool with VBFNLO as a function of higher-dimensional operators considered in our study are given in Fig.\ref{fig1}. The unitarity is safe in the region that is below the line for each coefficient. 

The recent experimental constraints on anomalous quartic gauge couplings $f_{T8}/\Lambda^{4}$ and $f_{T9}/\Lambda^{4}$ by CMS and ATLAS collaboration at 13 TeV center of mass energy and integrated luminosity of 137 and 139 fb$^{-1}$ at LHC are obtained via VBS and tri-boson production channels.
In these analysis, tri-boson production channel is treated as a Z boson production in association with two photons while VBS production channel is treated as electroweak (EW) production of a Z boson, a photon, and two forward jets or the electroweak production of two jets and a Z-boson pair. The decay of the Z-boson containing two charged leptons ($Z\to ll$) or two neutrinos ($Z\to \nu\nu$) leads to the different final states in both production channels. The obtained limit values without unitarity preserving on $f_{T8}/\Lambda^{4}$ and $f_{T9}/\Lambda^{4}$ by ATLAS and CMS collaborations are listed in Table \ref{tab6}. The best 95\% C.L. experimental limits on $f_{T8}/\Lambda^{4}$ and $f_{T9}/\Lambda^{4}$ obtained from analysis of the electroweak production of $Z(\nu\nu)\gamma$ in association with two jets by ATLAS collaboration at a center-of-mass energy of 13 TeV with an integrated luminosity 139 fb$^{-1}$ \cite{ATLAS:2022nru} are [-0.059; 0.059] TeV$^{-4}$ and [-0.13; 0.13] TeV$^{-4}$ , respectively.

\begin{table}
\caption{The 95\% C.L. experimental limits on the anomalous quartic gauge couplings $f_{T8}/\Lambda^{4}$ and $f_{T9}/\Lambda^{4}$ obtained without unitarity preserving from analysis of different production channels by ATLAS and CMS collaborations.}
\label{tab6}
\begin{ruledtabular}
\begin{tabular}{lccccr}
Channels&$f_{T8}/\Lambda^{4}$ (TeV$^{-4}$)&$f_{T9}/\Lambda^{4}$ (TeV$^{-4}$)&$\sqrt s$ (TeV)&L$_{int}$ (fb$^{-1}$)&Collaboration\\\hline
$Z(ll)\gamma\gamma$&[-1.06;1.10]&[-1.82;1.82]&13&137&CMS \cite{CMS:2021jji}\\
$Z(ll)Z(ll)jj$&[-0.43;0.43]&[-0.92;0.92]&13&137&CMS \cite{CMS:2020fqz}\\
$Z(ll)\gamma jj$&[-0.47;0.47]&[-0.91;0.91]&13&137&CMS \cite{CMS:2021gme}\\
$Z(\nu\bar\nu)\gamma jj$&[-0.059; 0.059]&[-0.13; 0.13]&13&139&ATLAS \cite{ATLAS:2022nru}\\
\end{tabular}
\end{ruledtabular}
\end{table}

\section{Signal and Background Simulation}
Before starting a detailed Monte Carlo simulation, it is important to start by examining the effects of anomalous quartic couplings on the cross section of tri-photon production to determine the analysis strategy.
\begin{figure}[h]
\includegraphics[scale=0.4]{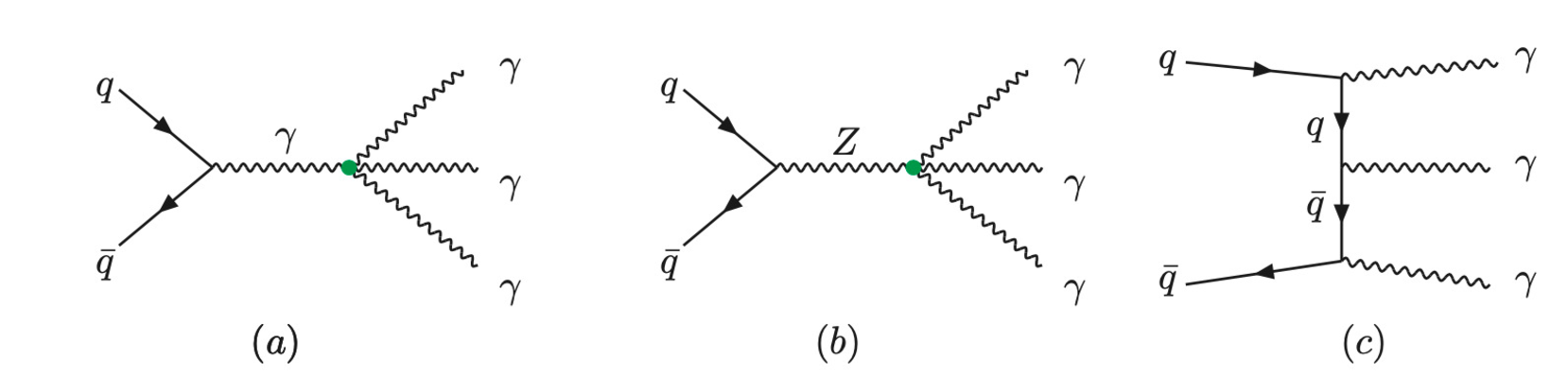} 
\caption{The Feynman diagrams contributing $\gamma\gamma\gamma$ production at hadron colliders  (a) and (b) through anomalous quartic vector-boson interactions with marked green (c) as well as SM contribution \label{fd}}
\end{figure}
The Feynman diagrams of $\gamma\gamma\gamma$ production, including anomalous quartic gauge boson couplings $f_{T,8}$ and $f_{T,9}$ (a)$\gamma\gamma\gamma\gamma$ and (b)$Z\gamma\gamma\gamma$ and also (c) SM contribution, are shown in Fig.~\ref{fd}. The total cross section corresponding to the sum of the three Feynman diagrams in this figure is a quadratic function for the case of only one non-zero anomalous quartic $f_{Ti}/\Lambda^{4}$ coupling at a time, as follows
\begin{eqnarray}\label{cs}
 \sigma_{tot}=\sigma_{SM}+(\frac{f_{Ti}}{\Lambda^{4}})^2\sigma_{\gamma}+(\frac{f_{Ti}}{\Lambda^{4}})^2\sigma_{Z}+(\frac{f_{Ti}}{\Lambda^{4}})^2\sigma_{\gamma Z}+\frac{f_{Ti}}{\Lambda^{4}}\sigma_{\gamma-SM}+\frac{f_{Ti}}{\Lambda^{4}}\sigma_{Z-SM}
\end{eqnarray}
where $\sigma_{SM}$ is the SM cross section of $\gamma\gamma\gamma$ production (Fig.2(c)); $\sigma_{\gamma}$, $\sigma_{Z}$ and $\sigma_{\gamma Z}$ are pure anomalous cross sections due to $\gamma\gamma\gamma\gamma$ (Fig.\ref{fd}(a)), $Z\gamma\gamma\gamma$ (Fig.\ref{fd}(b)) vertices and the interference between them, respectively; $\sigma_{\gamma-SM}$ and $\sigma_{Z-SM}$ indicate interference between SM processes and the anomalous contribution to the $\gamma\gamma\gamma$ production. Numerical calculations of the cross section of $pp\to \gamma\gamma\gamma$ are performed in {\sc MadGraph5\_aMC@NLO} \cite{Alwall:2014hca} by including effective vertices of $\gamma\gamma\gamma\gamma$ and $Z\gamma\gamma\gamma$ aQGCs based on the Universal {\sc FeynRules} Output (UFO) framework \cite{Alloul:2013bka,Degrande:2011ua}. 
In Fig.\ref{cs_without}, we show the contribution of each term in Eq.~(\ref{cs}) separately as a function of $f_{T8}/\Lambda^{4}$ and $f_{T9}/\Lambda^{4}$ assuming that only one anomalous quartic coupling is different from zero at a time, which gives the contributions of the $\gamma\gamma\gamma\gamma$  and $Z\gamma\gamma\gamma$ vertices separately. 
In order to imitate the performance of a typical electromagnetic calorimeter and make sure that cross sections are free of infrared divergences, we applied a cut on the transverse momentum of photons in the final state as a $p_T^{\gamma}>$ 25 GeV at the generator level.

The results of Fig.\ref{cs_without} found clear support for cross section of the three photon production are most sensitive to $f_{T8}/\Lambda^{4}$ coupling than to the $f_{T9}/\Lambda^{4}$ coupling. Further novel finding form  Fig.\ref{cs_without} is that the main contribution to the variation of the cross section with respect to both $f_{T8}/\Lambda^{4}$ and $f_{T9}/\Lambda^{4}$ comes from the $\gamma\gamma\gamma\gamma$ vertex and its interference with the SM rather than the $Z\gamma\gamma\gamma$ vertex.

\begin{figure}[htbp]
\includegraphics[scale=0.6]{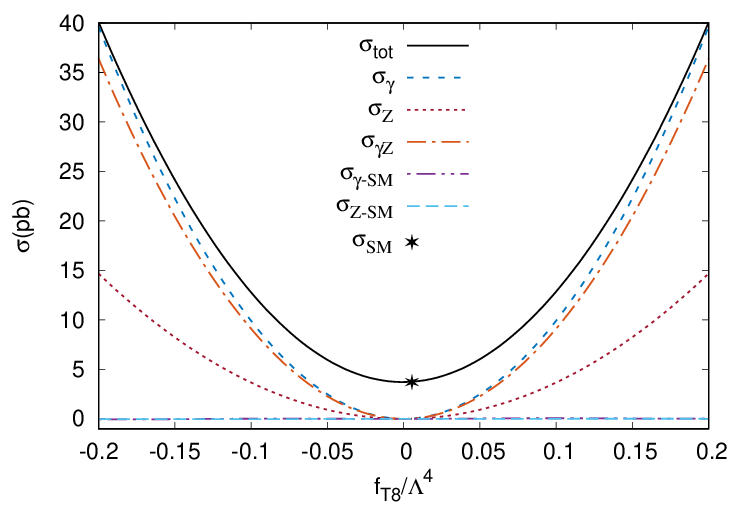}
\includegraphics[scale=0.6]{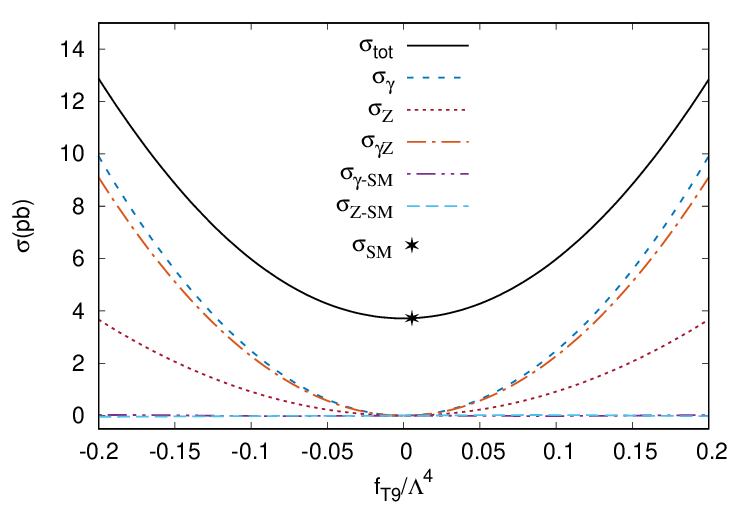}
\caption{The contribution of $\gamma\gamma\gamma\gamma$($\sigma_{\gamma}$), $Z\gamma\gamma\gamma$($\sigma_{Z}$) vertex and their interference with each other ($\sigma_{\gamma Z}$) and SM ($\sigma_{\gamma-Z}$ and $\sigma_{Z-SM}$) to the cross section for the process $pp \to \gamma \gamma\gamma$ as function of dimension-8 aQGCs $f_{T8}/\Lambda^4$ and $f_{T9}/\Lambda^4$ at $\sqrt s$=100 TeV without form factors applied.\label{cs_without}}
\end{figure}

We study dimension-8 operators via $pp\to\gamma\gamma\gamma$ signal process at FCC-hh since it is a clean probe for new physics. Although our analysis strategy relies mainly on the isolated three photons in the final state, the considered signal process suffers from the backgrounds from high-$p_{T}$ jets, which are misidentified as isolated photons even after imposing the tight identification and isolation requirements. Therefore following relevant backgrounds are considered; multi-jet (${j}{j}{j}$), photon + jet ($\gamma{j}{j}$), diphoton-jet ($\gamma\gamma{j}$). In order to perform detailed signal and background analysis, for $\gamma\gamma\gamma$ (SM), $\gamma\gamma{j}$, $\gamma{j}{j}$, and ${j}{j}{j}$  background processes and five different positive values of each $f_{T8}/\Lambda^{4}$ and $f_{T9}/\Lambda^{4}$ couplings for signals 600.000 (or 600k) events are generated within {\sc MadGraph5\_aMC@NLO} using NNPDF23LO PDF set \cite{Ball:2012cx}. Parton level events are then passed through to {\sc Pythia 8.20} \cite{Sjostrand:2014zea} for parton showering and hadronization. Detector response is included in then hadronized events by {\sc Delphes 3.4.2} \cite{deFavereau:2013fsa} software with card namely \verb|FCC-hh.tcl| where detector response in the form of resolution functions and efficiencies are parametrised. Jets are reconstructed by using clustered energy deposits with {\sc FastJet 3.3.2} \cite{Cacciari:2011ma} using anti-kt algorithm \cite{Cacciari:2008gp} where a cone radius is set as $\Delta R$ = 0.2 and $p_T^j>$ 20 GeV.
\begin{table}
\caption{The total cross sections for production of $\gamma\gamma\gamma$, $\gamma\gamma{j}$, $\gamma{j}{j}$, and ${j}{j}{j}$ processes at $\sqrt s$=14 TeV (with $p_T^{\gamma}>15$ GeV) and $\sqrt s$=100 TeV (with $p_T^{\gamma}>25$ GeV)  \label{crosssection}}
\begin{ruledtabular}
\begin{tabular}{lcc}
 Processes& \multicolumn{2}{c}{Total cross sections (pb)}  \\ \hline
&$\sqrt s$=14 TeV  & $\sqrt s$=100 TeV  \\\cline{2-3}
$pp\to\gamma\gamma\gamma$ & 4.349$\times10^{-2}$&5.741$\times10^{-2}$ \\
$pp\to\gamma\gamma{j}$& 5.405$\times10^{1}$&1.464$\times10^{2}$ \\
$pp\to\gamma{j}{j}$&5.148$\times10^{4}$ &2.181$\times10^{5}$\\
$pp\to{j}{j}{j}$&8.022$\times10^{7}$&1.239$\times10^{9}$  \\
\end{tabular}
\end{ruledtabular}
\end{table}

In Table \ref{crosssection}, to compare total cross section for production of the $\gamma\gamma\gamma$, $\gamma\gamma{j}$, $\gamma{j}{j}$, and ${j}{j}{j}$ background prosseses at HL-LHC ($\sqrt s$=14 TeV)  and FCC-hh( $\sqrt s$=100 TeV) are given. In the calculations of the cross sections at the Leading Order (LO)  in Table \ref{crosssection}, we apply generator level cuts; the rapidity of each photon $|\eta^{\gamma}|< 2.5$, separation between any two photons ,two jets, a photon and a jet in the ($\eta$-$\Phi$) plane $\Delta R(\gamma, \gamma)>0.4$, $\Delta R(\gamma, j)>0.4$,$\Delta R(j, j)>0.4$ where $\Delta R(\gamma(j)\gamma(j))=[\Delta \eta^2+\Delta\Phi^2]^{1/2}$. In addition, we have apply two types of cuts on the transverse momentum of each photon i.e., $p_{T}^{\gamma}>$ 15 GeV for HL-LHC and $p_{T}^{\gamma}>$25 GeV for FCC-hh. While the main contribution to the total cross section of the $pp\to\gamma\gamma\gamma$ process comes only from the  $qq\to\gamma\gamma\gamma$ subprocess as seen in the Feynman diagrams shown in Fig.\ref{fd}, the main contribution to the $pp\to\gamma\gamma j$ and $pp\to\gamma jj$ processes comes from the $qg\to\gamma\gamma q$ and $qg\to\gamma qg$ subprocesses, respectively. Besides the $qg\to ggq$ subprocess, the most significant contribution to the total cross section of the $pp\to jjj$ process comes from the $gg\to ggg$ subprocess. Therefore, the contributions from the gluon initial state subprocesses increase the cross sections of the $pp\to\gamma\gamma j$, $pp\to\gamma jj$ and $pp\to jjj$ processes to quite large values.

%Tri-photon production has already been studied at next-to-leading order (NLO) \cite{Bozzi:2011en,Mandal:2014vpa,ATLAS:2015rsn}, as well as at next-to-next-to-leading order (NNLO)  level  \cite{Chawdhry:2019bji,Kallweit:2020gcp,Abreu:2020cwb,Abreu:2021oya}. Since tri-photon production does not depend on the strong coupling at the leading order, there is only a very small dependence on the factorization scale at that order. When considering NLO prediction, it becomes sensitive to the gluon distribution and leading to significant K-factors ($\sim 2 - 2.5$) when going from LO to NLO \cite{Campbell:2014yka}.

 NLO predictions for signal and background processes in the short-distance collisions at a hadron collider rely on perturbative QCD. However, leading order (LO) predictions in QCD suffer from a dependence on non-physical renormalization and factorization scales, which is further strengthened by the increasing jet multiplicity.  Thus, next-to-leading order (NLO) results generally reduce this dependence dramatically. Since tri-photon production does not depend on the strong coupling at the leading order, there is only a very small dependence on the factorization scale at that order. 
 Tri-photon production has already been studied at NLO \cite{Bozzi:2011en,Mandal:2014vpa,ATLAS:2015rsn}, as well as at next-to-next-to-leading order (NNLO)  level  \cite{Chawdhry:2019bji,Kallweit:2020gcp,Abreu:2020cwb}. When considering NLO prediction, it becomes sensitive to the gluon distribution and leading to significant K-factors ($\sim 2 - 2.5$) when going from LO to NLO \cite{Campbell:2014yka}. NLO and NNLO predictions for the background $pp\to\gamma\gamma j$ \cite{Gehrmann:2013aga,Chawdhry:2021hkp}, $pp\to\gamma jj$ \cite{ATLAS:2019iaa} and $pp\to jjj$ \cite{ATLAS:2015yaa, Abreu:2021oya} processes are also studied in detail. Finally, effects of the NLO prediction on the cross sections might be significant, therefore affect the background events estimation. However, this effect was considered as a part of the systematic uncertainty in the evaluation of statistical significance for our study.
 
The probability of misidentifying a jet as a photon depends on the interaction between the jets and the detector. Presently, the probability of a hard scattering jet being inaccurately reconstructed as an isolated photon is minimal due to the exceptional angular resolution of existing calorimeters in LHC detectors. It is anticipated that the granularity of detectors in upcoming hadron-hadron colliders, which will employ detector parameters, will be 2-4 times more precise than those in current LHC detectors. Therefore we applied a constant jet-to-photon fake rate of $10^{-3}$ to each jet in the $\gamma\gamma{j}$, $\gamma{j}{j}$, and ${j}{j}{j}$ background data samples.
 
\section{Event Selection and analysis with boosted DECISION TREES}
Machine learning algorithms are increasingly popular in particle physics analysis due to their ability to solve particle classification and event selection problems and their usability for reconstruction, a regression task. Among these algorithms, the use of multivariate techniques in analyzing data from colliders has come to the forefront. The use of boosted decision trees (BDTs) in multivariate techniques offers new windows not only for the analysis of big data from particle colliders but also for the possibility of increasing signal significance by relying on the algorithm to separate the signal from the relevant background to identify new physics signal events. Therefore, we use the Boosted Decision Trees (BDT) method in the toolkit \cite{Roe:2004na,Carleo:2019ptp} for multivariate analysis (TMVA) \cite{Hocker:2007ht,Therhaag:2009dp} for the analysis of the tri-photon process at the FCC-hh collider to obtain sensitivity on the dimension-8 effective operators.

Our multivariate analysis utilizes the BDT machine learning technique with 850 trees, a maximum depth of 3, minimum events at each final leaf of 2.5, a number of iterations to find the best split of 20, and a learning rate of 0.5. We use adaptive boosting (AdaBoost) mechanisms for training that signal events from the training sample that end up in a background node. We merge data featuring selected kinematic variables for both signal and background events into a unified data file. 50\% of the combined data is designated for training, while the remaining 50\% is reserved for testing.

\begin{table}
\caption{The list of selected kinematical and reconstructed variables to be used in BDT.   \label{BDT}}
\begin{ruledtabular}
\begin{tabular}{lll}
Variable & ~~~~~~~~~~~~Definition \\ \hline 
%$N_{b}$&Number of b-tagged jets in the event \\
%$N_{l}$&Number of leptons in the event \\
$p_{T}^{\gamma_1}$&Transverse momentum of the leading photon \\
$p_{T}^{\gamma_2}$&Transverse momentum of the second-highest-$p_{T}$ photon \\
$p_{T}^{\gamma_3}$&Transverse momentum of the third-highest-$p_{T}$ photon \\
%$p_{T}^{b}$&Transverse momentum of the leading b-jet \\
$\eta^{\gamma_1}$&Pseudo-rapidity of the leading photon \\
$\eta^{\gamma_2}$&Pseudo-rapidity of the second-highest-$p_{T}$ photon \\
$\eta^{\gamma_3}$&Pseudo-rapidity of the third-highest-$p_{T}$ photon \\
$\Delta|\phi^{\gamma_1\gamma_2}|$& Difference in azimuthal angle between the leading and second-highest-$p_{T}$ photons \\
$\Delta|\phi^{\gamma_1\gamma_3}|$&Difference in azimuthal angle between leading and the third-highest-$p_{T}$ photon \\
$\Delta|\phi^{\gamma_2\gamma_3}|$&Difference in azimuthal angle between second-highest-$p_{T}$ photon and the third-highest-$p_{T}$ photon \\
$\Delta|\eta^{\gamma_1\gamma_2}|$& Difference in pseudo-rapidity between the  leading and the second-highest-$p_{T}$ photons \\
$\Delta|\eta^{\gamma_1\gamma_3}|$&Difference in pseudo-rapidity between the  leading and the third-highest-$p_{T}$ photon \\
$\Delta|\eta^{\gamma_2\gamma_3}|$&Difference in pseudo-rapidity between the second-highest-$p_{T}$ and the third-highest-$p_{T}$ photon \\
$\Delta R (\gamma_1,\gamma_2)$&Distance between leading photon and sub-leading photon in $\eta$-$\phi$ plane \\
$\Delta R (\gamma_2,\gamma_3)$&Distance between sub-leading photon and third photon  in $\eta$-$\phi$ plane \\
$\Delta R (\gamma_1,\gamma_3)$&Distance between leading photon and third photon in $\eta$-$\phi$ plane \\
$N_{j}$&Number of jets in the event \\
$m_{\gamma_1\gamma_2\gamma_3}$&Invariant mass of reconstructed three photon system\\
\end{tabular}
\end{ruledtabular}
\end{table}
Event selection starts by requiring the presence of at least three photons in the signal and SM background processes, with generator level cuts on the final state particles. In order to make the most effective use of the kinematic differences between the signal and background processes, a Boosted Decision Tree (BDT) is trained using most of the available kinematic information from the event, given in Table \ref{BDT}. Input variables to a multivariate BDT discriminant can be classified as the transverse momentum and pseudo-rapidity of three-$p_T$ ordered photons ($\gamma_1$,$\gamma_2$ and $\gamma_3$), the difference in the azimuthal angle and in pseudo-rapidity between pairs of photons, the distance between pairs of photons in the $\eta$-$\phi$ plane and the invariant mass of three photon system ($m_{\gamma_1\gamma_2\gamma_3}$). Since we try to well separate our signal from SM backgrounds with jets in the final state, the number of jets is considered in the BDT input variable list. However, the kinematic variables of the jets are not included. Among the multivariate BDT discriminant variables listed in Table \ref{BDT}, the invariant mass of the tri-photon system is selected as a target variable in BDT analysis.

\begin{figure}[htbp]
\includegraphics[scale=0.37]{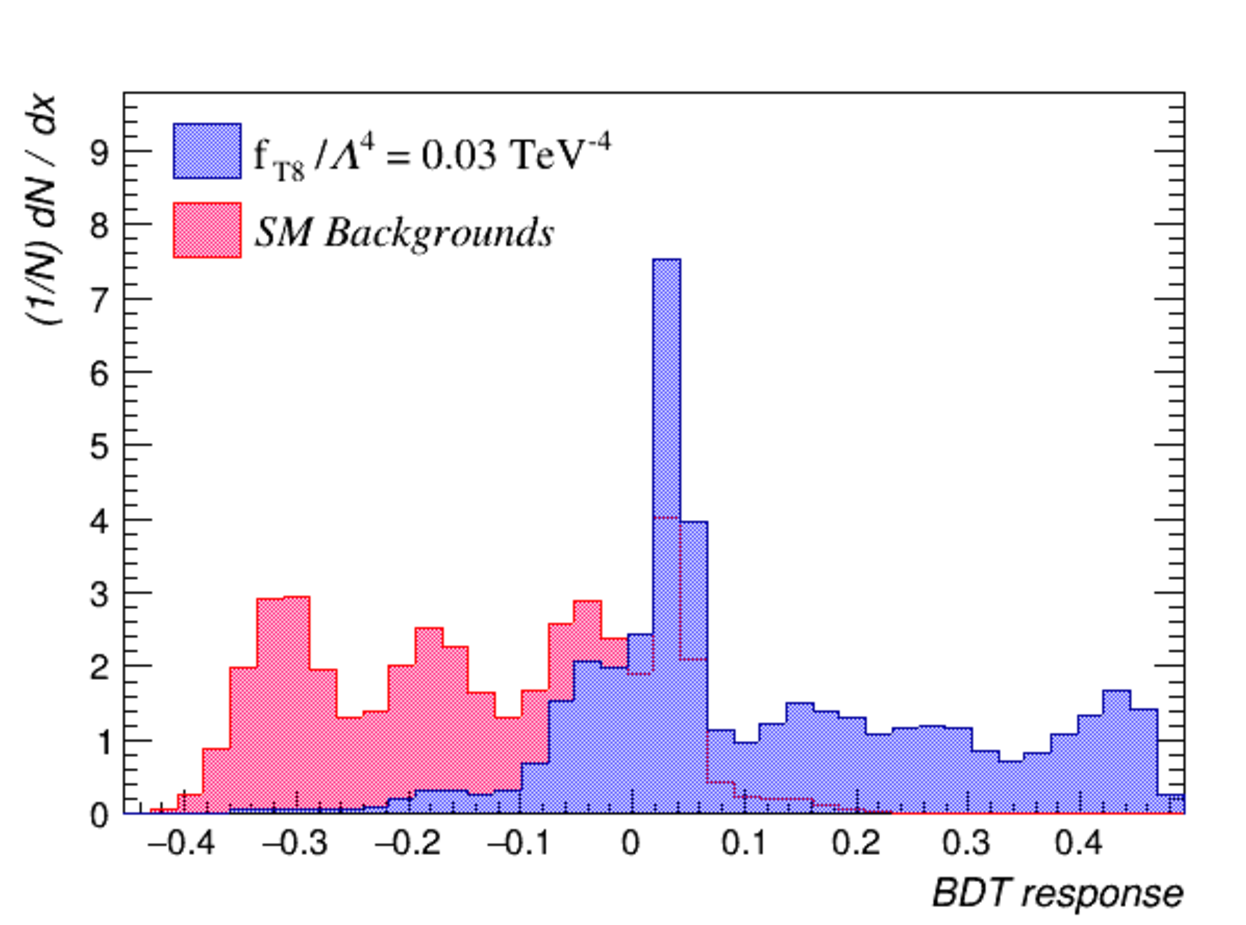}
\includegraphics[scale=0.37]{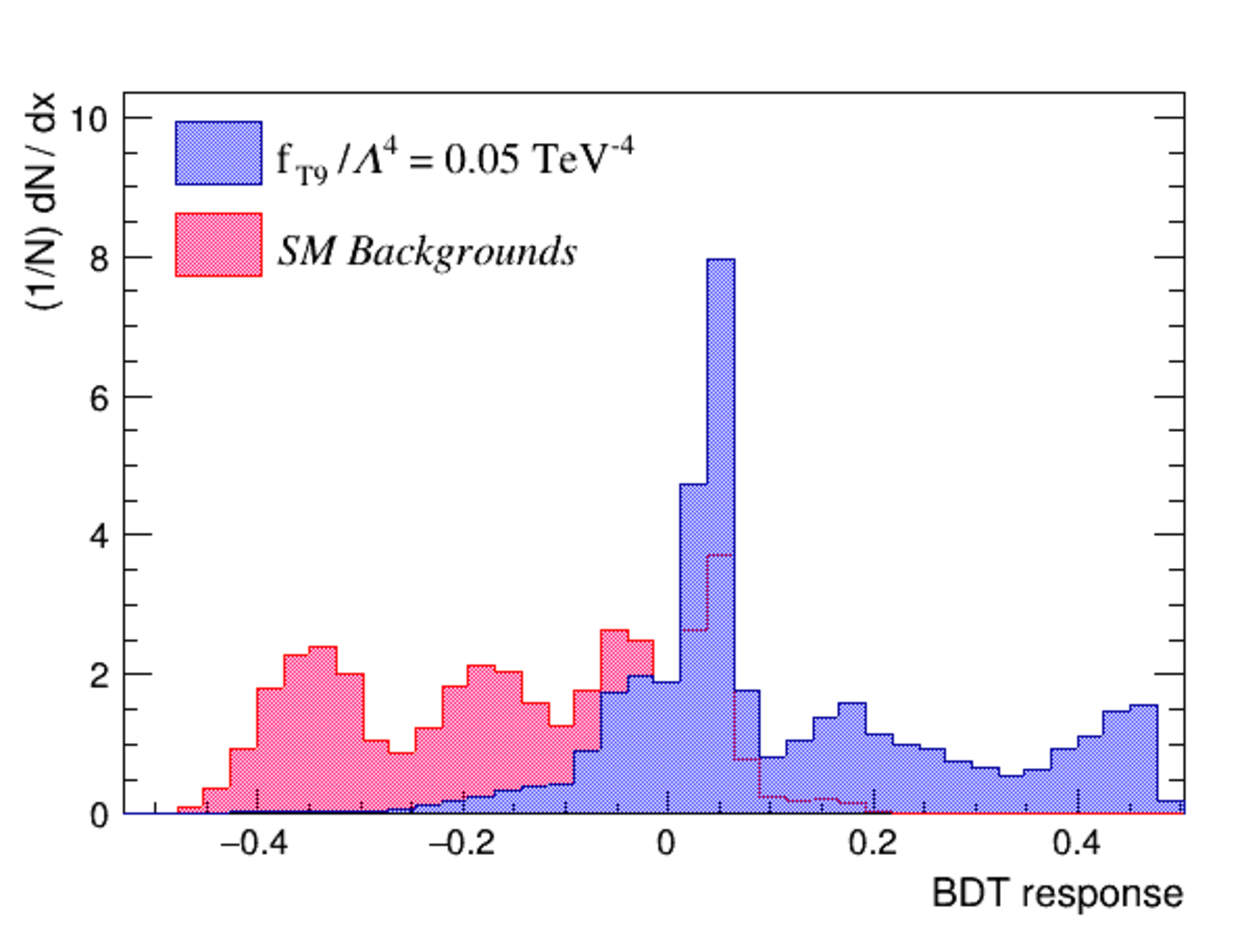}\\
\includegraphics[scale=0.39]{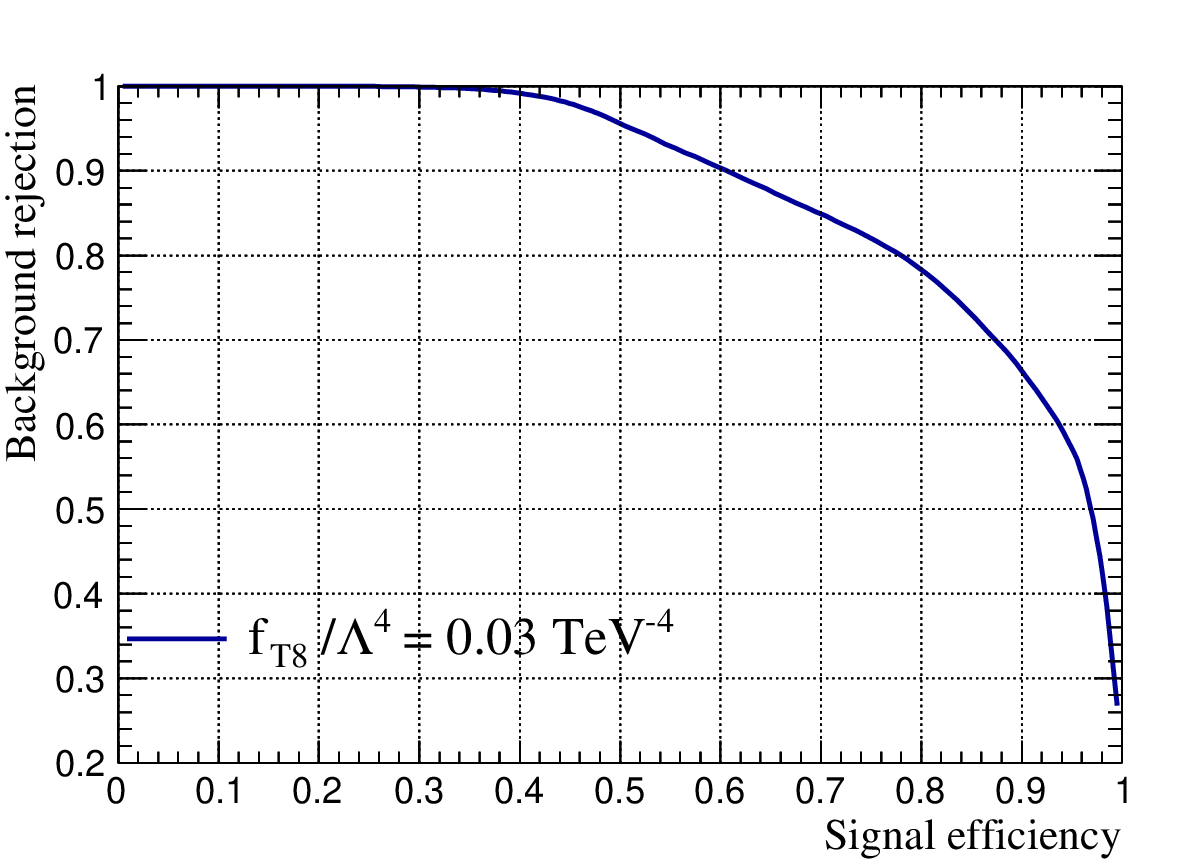}
\includegraphics[scale=0.39]{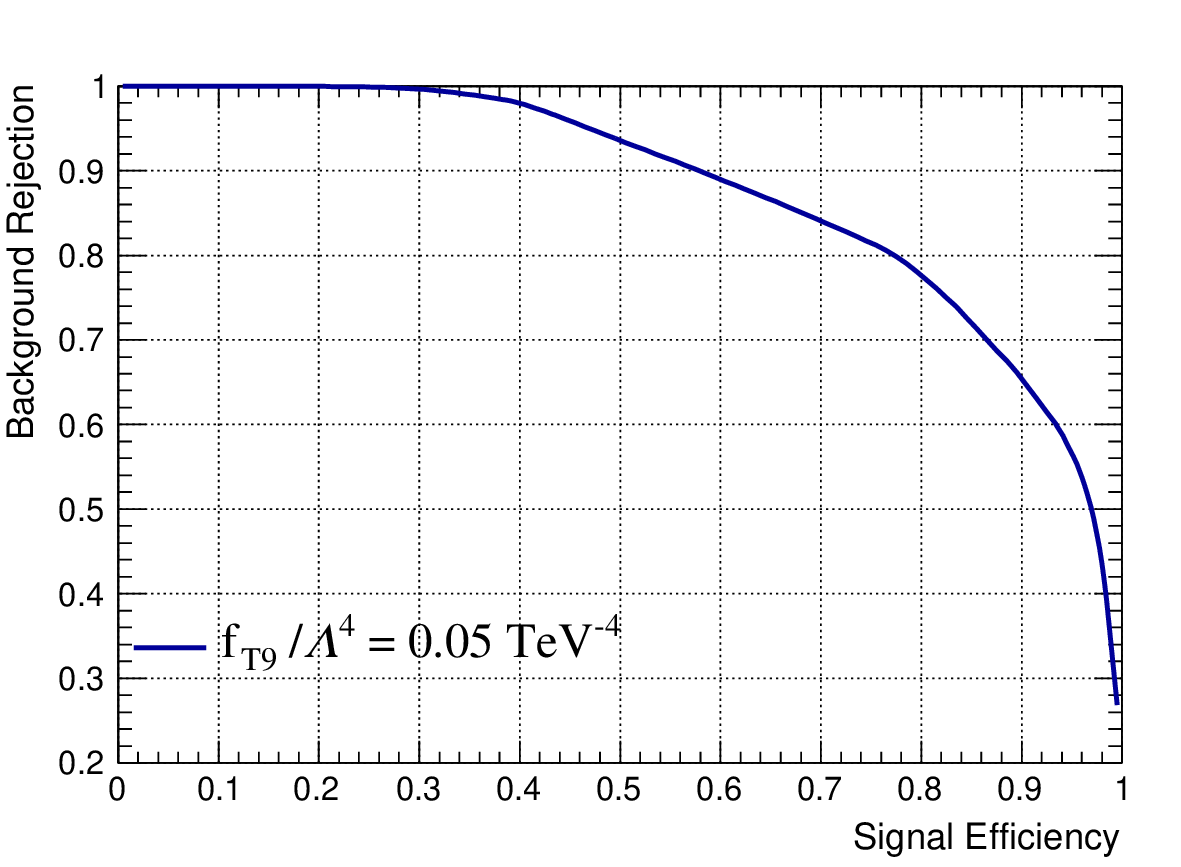}
\caption{The distribution of the BDT response (on the first row) and Receiver Operating Characteristic (ROC) curve of the BDT (on the second row) for signal ($f_{T8}/\Lambda^{4}$=0.03 TeV$^{-4}$ and $f_{T9}/\Lambda^{4}$=0.05 TeV$^{-4}$) and all relevant backgrounds. \label{BDT_output}}
\end{figure}

Fig.\ref{BDT_output} shows the BDT classifier response for the considered signal and total background events. The first row in Fig.\ref{BDT_output} corresponds to the BDT response, while the second row shows the signal efficiency as a function of background rejection, the so-called Receiver Operating Characteristic (ROC) curve. The left column in Fig.\ref{BDT_output} corresponds to the BDT response and ROC curve for the signal with $f_{T8}/\Lambda^{4}$=0.03 TeV$^{-4}$ and all the other overwhelming backgrounds for trained samples, while the right column is for the signal with $f_{T9}/\Lambda^{4}$=0.05 TeV$^{-4}$ value.
\begin{figure}[htbp]
\includegraphics[scale=0.4]{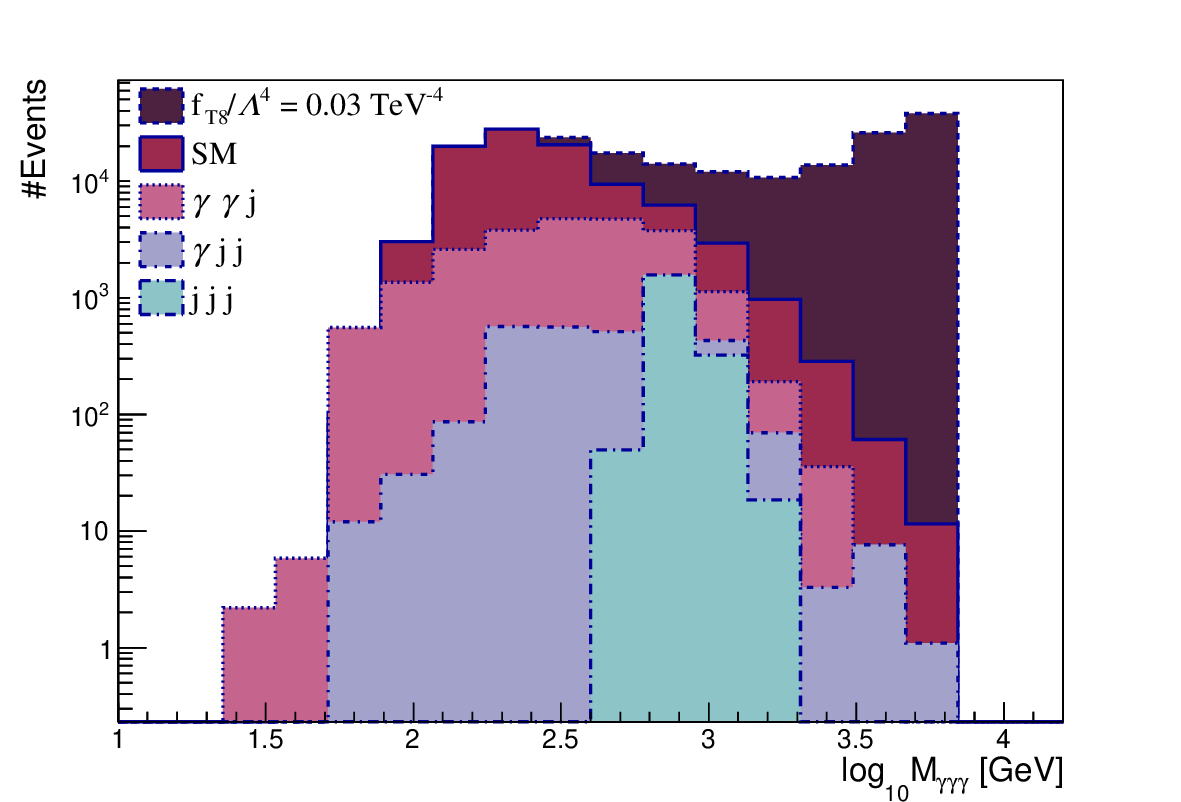}
\includegraphics[scale=0.4]{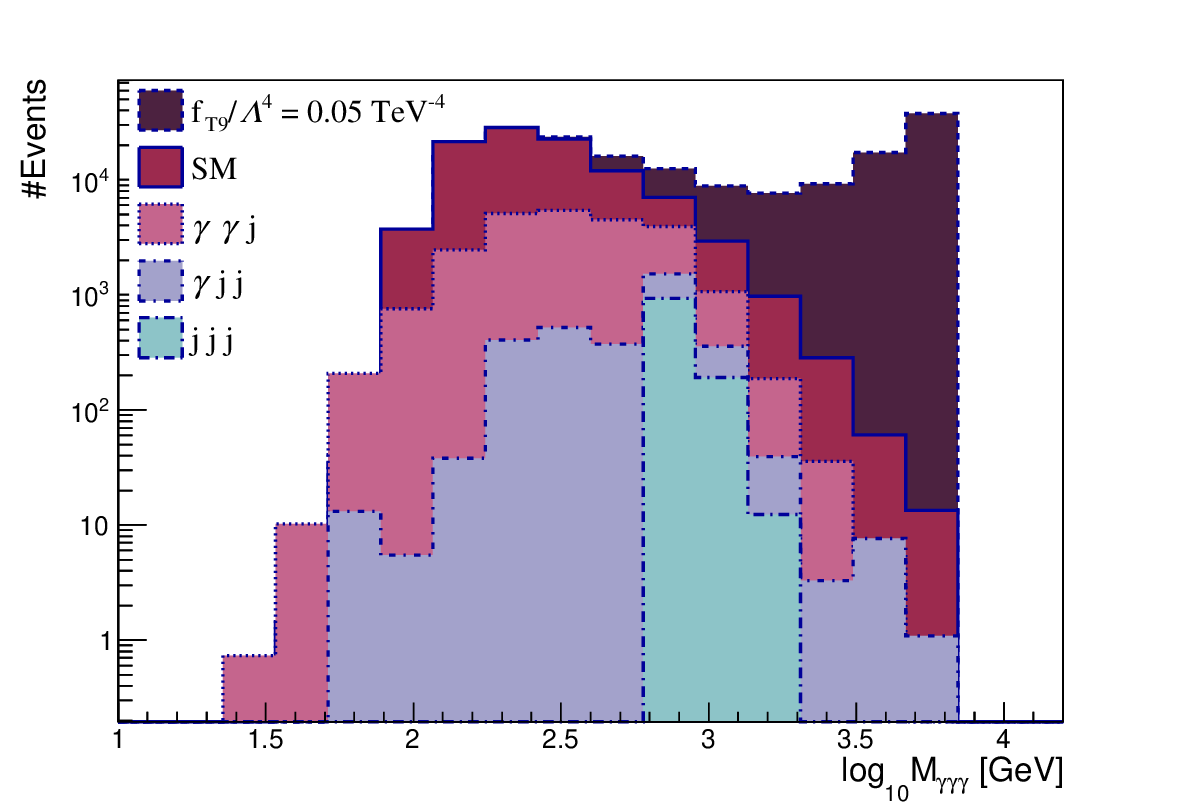}
\caption{The normalized distribution of invariant mass of reconstructed tri-photon system ($m_{\gamma\gamma\gamma}$) for $pp\to\gamma\gamma\gamma$ signal process with $f_{T8}/\Lambda^{4}$=0.03 TeV$^{-4}$ ($f_{T9}/\Lambda^{4}$=0.05 TeV$^{-4}$) aQGC values and all relevant backgrounds at FCC-hh with L$_{int}$=30 ab$^{-1}$. \label{maaa}}
\end{figure}

From this figure, it is evident that the BDT performs well in separating the signal from the backgrounds. We consider 70\% signal efficiency in the determination of an optimal cut on reconstructed BDT distributions. Since each signal, which depends on the different aQGC values of $f_{T8}/\Lambda^{4}$ and $f_{T9}/\Lambda^{4}$ couplings, is trained individually in the BDT analysis, we obtain a different BDT score for each signal and relevant SM backgrounds. Therefore, we apply the optimal BDT score obtained from the BDT response for each value of $f_{T8}/\Lambda^{4}$ and $f_{T9}/\Lambda^{4}$ aQGCs to the signal as well as relevant background processes to obtain the invariant mass distributions of three photon systems. 
 
The presence of effective operators results in a violation of unitarity at high energies; hence, it is necessary to restrict the application of parameterizing deviations from the SM predictions in terms of effective operator approximations to energy regimes where this violation does not occur. The method employed to avoid unitarity violation in this analysis is commonly known as the clipping method \cite{ATLAS:2022nru,CMS:2020gfh}, which involves suppressing any effective field theory (EFT) contribution beyond an energy scale $\Lambda_{FF}$. The energy scale, $\Lambda_{FF}$ for each aQGC parameter is calculated using VBFNLO after applying appropriate aQGC convention factors.

Then, the threshold of the clipping energy value for each aQGC parameter is applied to the invariant mass of $m_{\gamma\gamma\gamma}$. In Fig.\ref{maaa}, the normalized invariant mass distribution of three photons system for the signal with $f_{T8}/\Lambda^{4}$=0.03 TeV$^{-4}$ ($f_{T9}/\Lambda^{4}$=0.05 TeV$^{-4}$) benchmark points and relevant backgrounds are shown on the left (on the right). These figures are normalized to the cross-section of each process times the integrated luminosity, $L_{int}$ = 30 ab$^{-1}$. Only one coupling ($f_{T8}/\Lambda^{4}$ or $f_{T9}/\Lambda^{4}$) at a time is varied from its SM value.

\section{Limits on anomalous quartic gauge couplings}
In order to calculate the median expected significance for discovery and exclusion of the $f_{T8}/\Lambda^{4}$ and $f_{T9}/\Lambda^{4}$ couplings, denoted as $\mathcal{SS}_\text{disc}$ and $\mathcal{SS}_\text{excl}$, respectively, we use the expressions presented in ref. \cite{Cowan:2010js} 
 
\begin{eqnarray}
\mathcal{SS}_\text{disc}&=&
  \sqrt{2\left[(S+B)\ln\left(\frac{(S+B)(1+\delta^2 B)}{B+\delta^2 B(S+B)}\right) -
  \frac{1}{\delta^2 }\ln\left(1+\delta^2\frac{S}{1+\delta^2 B}\right)\right]} \\
    \mathcal{SS}_\text{excl} &=&\sqrt{2\left[S-B\ln\left(\frac{B+S+x}{2B}\right) 
  - \frac{1}{\delta^2 }\ln\left(\frac{B-S+x}{2B}\right)\right] -
  \left(B+S-x\right)\left(1+\frac{1}{\delta^2 B}\right)}.
\end{eqnarray}

where $\quad x=\sqrt{(S+B)^2- 4 \delta^2 S B^2/(1+\delta^2 B)}$;  $S$ and $B$ are the number of events obtained by integrating the normalized invariant mass distributions of the three photon system of the signal and total SM background, respectively, and $\delta$ is the systematic uncertainty. In the limit of $\delta \to 0$, these expressions can be simplified as
\begin{eqnarray}
\mathcal{SS}_\text{disc} &=& \sqrt{2[(S+B)\ln(1+S/B)-S]}\\
\mathcal{SS}_\text{excl} &=& \sqrt{2[S-B\ln(1+S/B)]} 
\end{eqnarray}
Regions with a $\mathcal{SS}_\text{disc}$ $\geqslant$ 5 (3) $\sigma$ are categorized as discoverable regions, while regions with a $\mathcal{SS}_\text{excl}$  $\leqslant$1.645 are considered regions that can be excluded at a 95\% confidence level. The source of systematic uncertainties are mainly based on the cross section measurements of $pp\to\gamma\gamma\gamma$ process with LO or NLO predictions and higher order EW corrections, the uncertainty in integrated luminosity as well as  jets misidentified as photons. In our study, we focus on LO predictions but do not investigate the impact and validity of these higher-order corrections on the signal and SM background processes. On the other hand, the total systematic uncertainty in the isolated three photon analysis with a 8 TeV centre-of-mass energy  and an integrated luminosity of 20.2 fb$^{-1}$ of the ATLAS Collaboration \cite{ATLAS:2017lpx} was determined to be approximately 13\%. Considering the timescale of the FCC-hh, overall uncertainty on the three photon production cross section can be improved with the theoretical predictions as well as new techniques to extract the luminosity can improve. Since the main purpose of this study is not to discuss sources of the systematic uncertainty in detail but to investigate the overall effects of the systematic uncertainty on the limits values of aQGC, we have considered \%10  overall systematic uncertainty in order to present optimistic results.

\begin{figure}[htbp]
\includegraphics[scale=0.4]{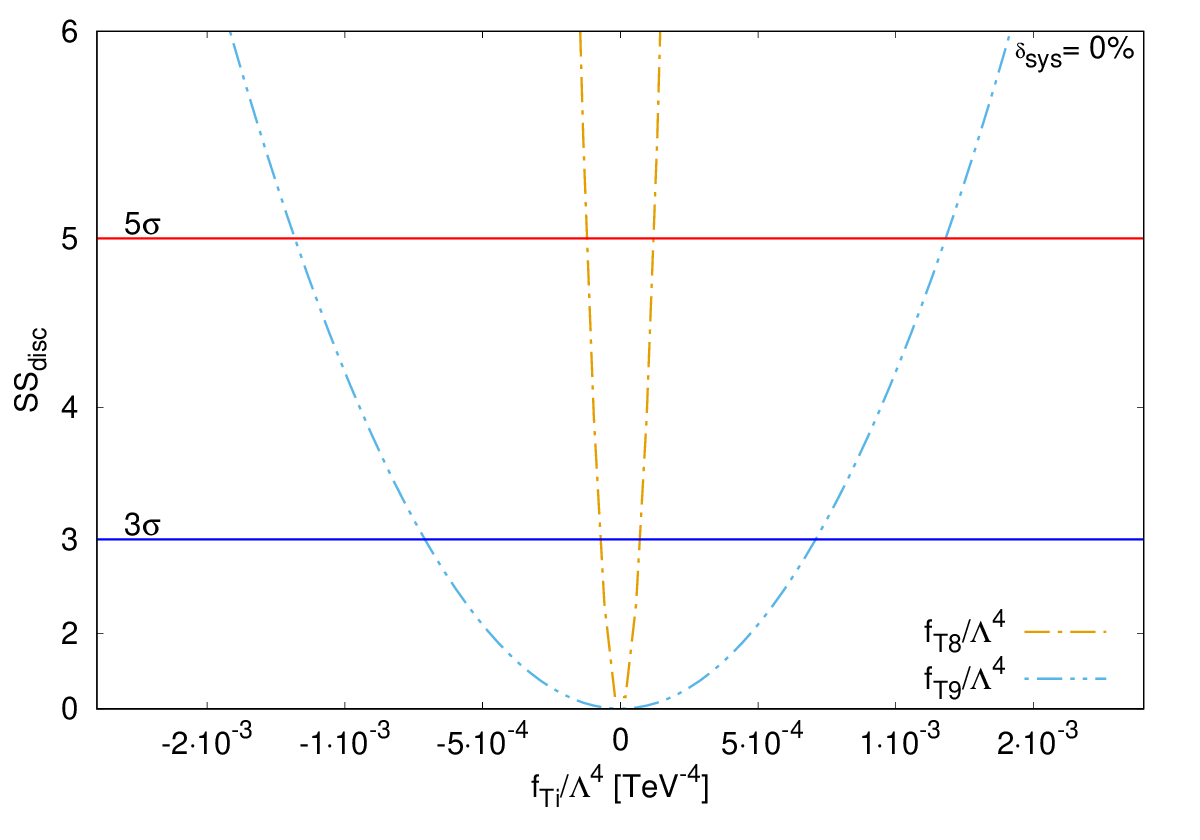}\includegraphics[scale=0.4]{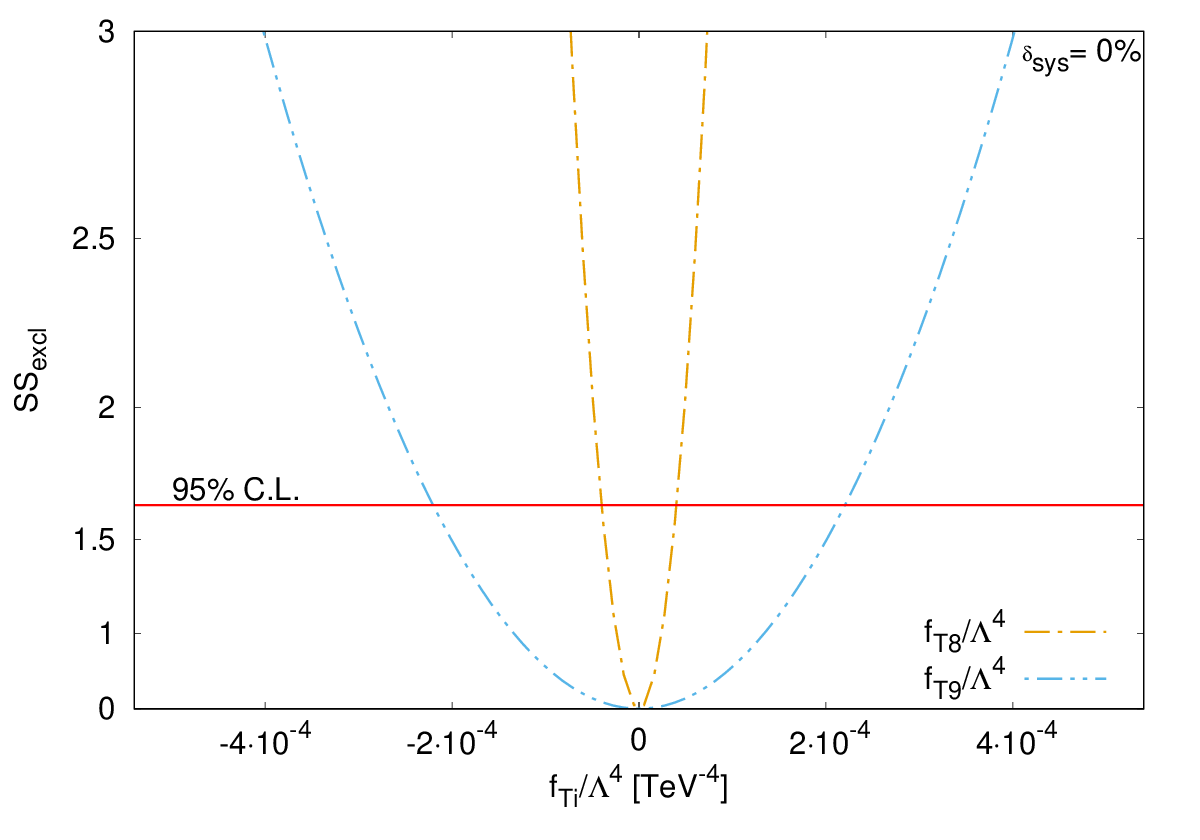}\\
\includegraphics[scale=0.4]{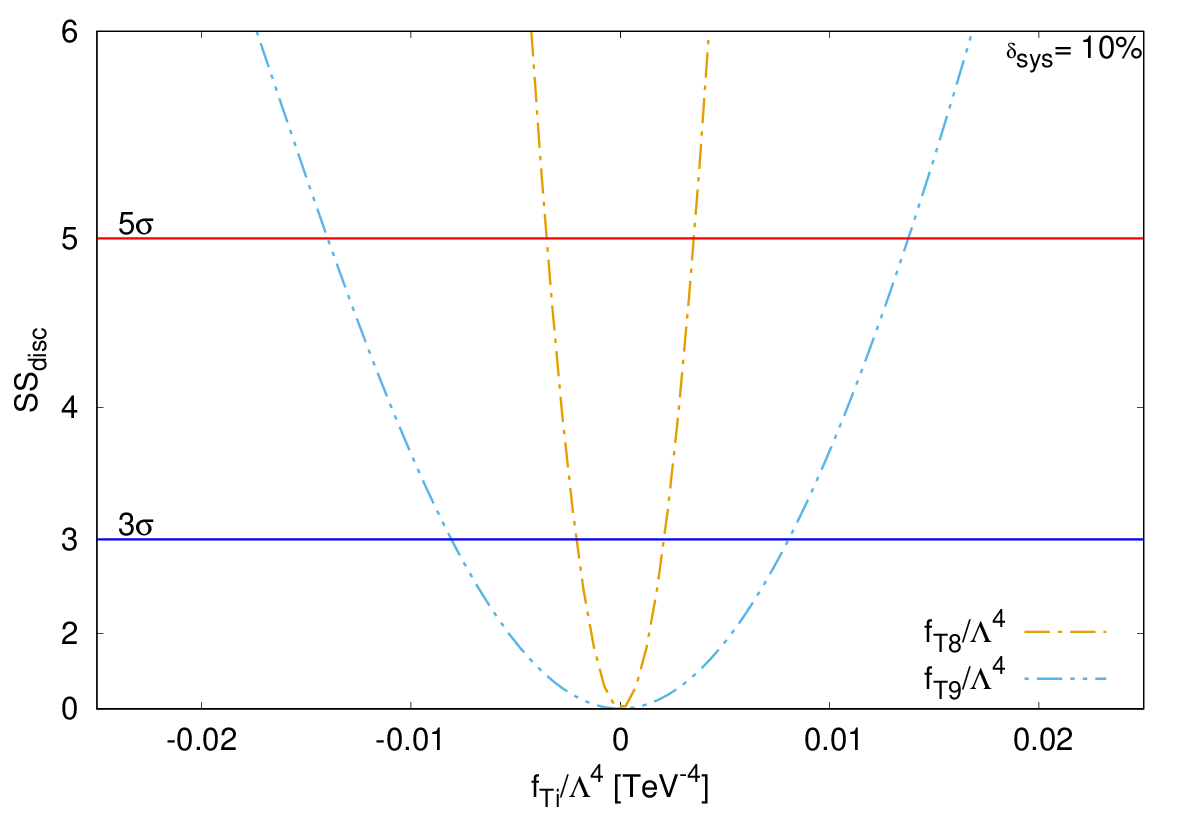}\includegraphics[scale=0.4]{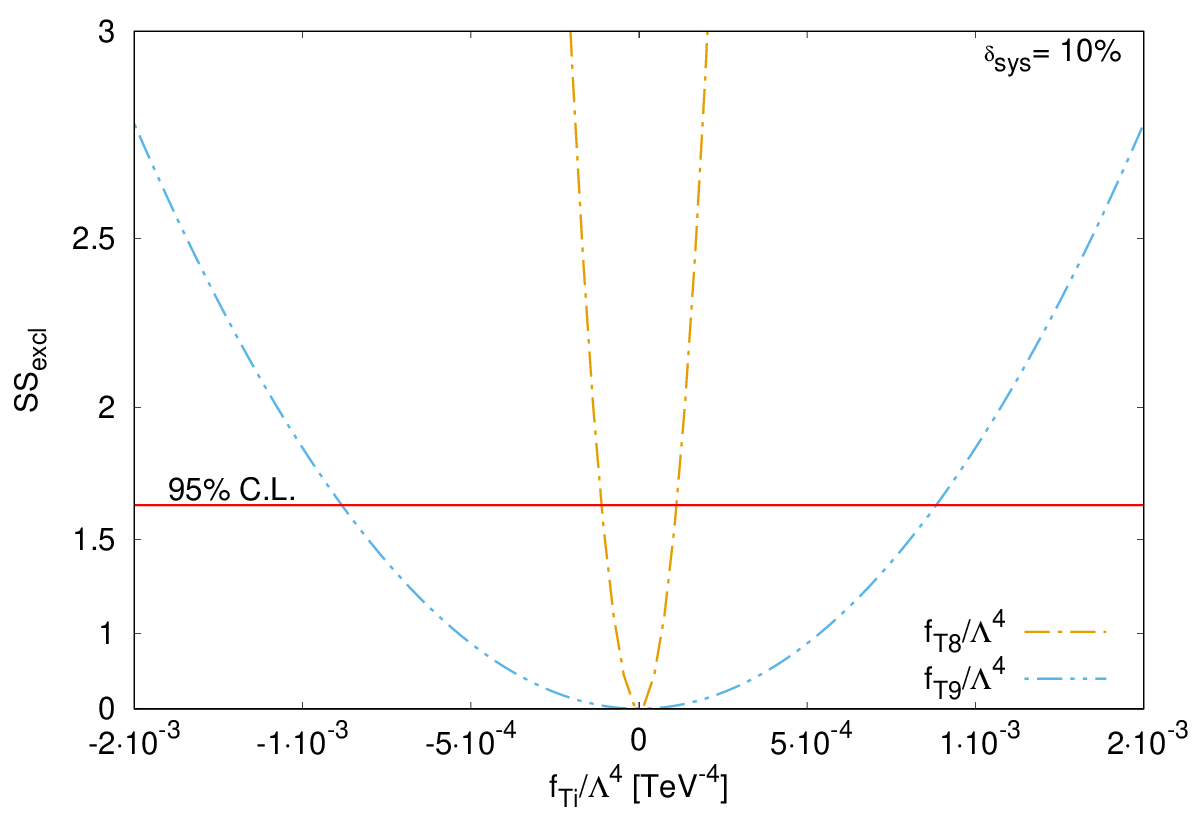}

\caption{The $\mathcal{SS}_\text{disc}$ and $\mathcal{SS}_\text{excl}$ as a function of the $f_{T8}/\Lambda^{4}$ and $f_{T8}/\Lambda^{4}$ after applying an optimum BDT cut value with and without systematic uncertainty at FCC-hh with an integrated luminosity L$_{int}$=30 ab$^{-1}$.  \label{SS}}
\end{figure}

The $\mathcal{SS}_\text{disc}$ and $\mathcal{SS}_\text{excl}$ as a function of aQGCs $f_{T8}/\Lambda^{4}$ (on the left panel) and $f_{T9}/\Lambda^{4}$  couplings for L$_{int}$=30 ab$^{-1}$ without (on the first row) and with $\delta \to 10\%$  (on the second row) systematic uncertainty are shown in Fig. \ref{SS}. In this figure, only one coupling at a time is varied from its SM value. From these figures, limits on dimension-eight anomalous quartic gauge couplings $f_{T8}/\Lambda^{4}$ and $f_{T9}/\Lambda^{4}$ can be inferred from the intersection of curves with horizontal red and blue lines representing by $3\sigma$ and $5\sigma$ levels in the left panel and with a red line representing 95\% C.L. in the right panel. Our obtained limits for $3\sigma$, $5\sigma$ and 95\% C.L., as well as without and with $\delta \to 10\%$  systematic uncertainties on dimension-eight anomalous quartic gauge couplings $f_{T8}/\Lambda^{4}$ and $f_{T9}/\Lambda^{4}$ with an integrated luminosity L$_{int}$=30 ab$^{-1}$ are listed in Tables \ref{limitsft8} and \ref{limitsft9}, respectively. 
The ATLAS collaboration recently reported best limits on aQGCs $f_{T8}/\Lambda^{4}$ and $f_{T9}/\Lambda^{4}$ based on the measured cross section of electroweak production of $Z(\nu\bar \nu)\gamma$ in association with two jets where they also applied the clipping technique. At 95\% C.L., current limits obtained with an integrated luminosity of 139 fb$^{-1}$ at $\sqrt s$=13 TeV are $[-5.2; 5.2]\times 10^{-1}$ TeV$^{-4}$ (with $\Lambda_{FF}$=1.7 TeV) and $[-7.9;7.9]\times 10^{-1}$ TeV$^{-4}$ (with $\Lambda_{FF}$=1.9 TeV) for $f_{T8}/\Lambda^{4}$ and $f_{T9}/\Lambda^{4}$, respectively.

As seen from the Table \ref{limitsft8} (\ref{limitsft9}), our limits on anomalous quartic gauge couplings $f_{T8}/\Lambda^{4}$ ($f_{T9}/\Lambda^{4}$) without systematic error at $5\sigma$ and 95\% C.L. are [-7.31;7.31]$\times 10^{-5}$ ([-4.59;4.59]$\times 10^{-4}$) TeV$^{-4}$ and [-4.01;4.01]$\times 10^{-5}$ ([-2.20;2.20]$\times 10^{-4}$) TeV$^{-4}$, respectively. Comparing our limits with those recently reported by ATLAS collaboration, we obtain a result more than three orders of magnitude better. Even considering 10\% systematic error, our results at $5\sigma$ and 95\% C.L. are still two orders of magnitude better.

Our results can also compare with the limits on aQGC $f_{T8}/\Lambda^{4}$ and $f_{T9}/\Lambda^{4}$ obtained from phenomenological studies via different production mechanism at future collider options.  In Ref. \cite{Yang:2021pcf}, the limit obtained by avoiding unitarity violation on $f_{T9}/\Lambda^{4}$ couplings via the $pp\to jjl^+l^-\gamma$ process for a center of mass energy of 14 TeV and an integrated luminosity of 3 ab$^{-1}$ is [-0.15;0.15] TeV$^{-4}$. In Ref. \cite{Degrande:2013yda}, obtained 95\% C.L. limits on aQGC $f_{T8}/\Lambda^{4}$ and $f_{T9}/\Lambda^{4}$ via $pp\to ZZ + 2j \to 4l + 2j$ processes  are  [-1.0;1.0] and [2.5;2.5] TeV$^{-4}$ at a centre-of-mass energy of 33 TeV with an integrated luminosity 3 ab$^{-1}$, respectively. In another study at hadron-hadron colliders \cite{Senol:2021wza}, sensitivity on aQGC $f_{T8}/\Lambda^{4}$ and $f_{T9}/\Lambda^{4}$  with MC simulation based on  $pp\to Z\gamma\gamma$ process with 100 TeV center of mass energies and an integrated luminosity 30 ab$^{-1}$ is [-1.16;0.54$]\times10^{-3}$ and [1.26;1.09]$\times10^{-3}$ TeV$^{-4}$, respectively. In Ref \cite{Ari:2021rmx}, the $ZZ\gamma\gamma$ and $Z\gamma\gamma\gamma$ quartic vertices are studied through the process $e^+e^-\to Z\gamma\gamma$ at the proposed 3 TeV $e^+e^-$ colliders and obtained limits on aQGC $f_{T8}/\Lambda^{4}$ ($f_{T9}/\Lambda^{4}$) of is [-1.78;1.85]$\times10^{-2}$ ([-3.83;3.48]$\times10^{-2}$) TeV$^{-4}$ with an integrated luminosity 5 ab$^{-1}$. The tri-photon production is also phenomenologically studied for different center of mass energy and integrated luminosities of proposed future muon collider \cite{Yang:2020rjt}. At 3 TeV center of mass energy and 1 ab$^{-1}$ integrated luminosity, the expected constraints on the $f_{T8}/\Lambda^{4}$ ($f_{T9}/\Lambda^{4}$) of is [-1.74;0.42]$\times10^{-2}$ ([-4.50;0.63]$\times10^{-2}$) TeV$^{-4}$. Our results considering 10\% systematic error at $5\sigma$ and 95\% C.L. are still better or comparable with the results of phenomenological studies at future lepton and hadron colliders.

\begin{table}
\caption{The limits on the $f_{T8}/\Lambda^4$ [TeV$^{-4}$] at $3\sigma$, $5\sigma$ and \% 95 C.L. without and with $\delta \to 10\%$ systematic uncertainties for $L_{int}$=30 ab$^{-1}$ at FCC-hh. \label{limitsft8}}
\begin{ruledtabular}
\begin{tabular}{lccc}
$\delta_{sys}$ & $3\sigma$  & $5\sigma$& 95\% C.L.  \\ \hline
0 & [-7.31; 7.31]$\times 10^{-5}$ & [-1.21;1.21]$\times 10^{-4}$ & [-4.01;4.01]$\times 10^{-5}$\\
10\%&[ -2.98; 2.98]$\times 10^{-3}$ &[-3.51;3.51]$\times 10^{-3}$& [-1.11;1.11]$\times 10^{-4}$ \\
\end{tabular}
\end{ruledtabular}
\end{table}

\begin{table}
\caption{The limits on the $f_{T9}/\Lambda^4$ [TeV$^{-4}$] at $3\sigma$, $5\sigma$ and \% 95 C.L. without and with $\delta \to 10\%$ systematic uncertainties for $L_{int}$=30 ab$^{-1}$ at FCC-hh.  \label{limitsft9}}
\begin{ruledtabular}
\begin{tabular}{lccc}
$\delta_{sys}$ & $3\sigma$  & $5\sigma$& 95\% C.L.  \\ \hline
0 & [-4.59; 4.59]$\times 10^{-4}$ & [-7.65;7.65]$\times 10^{-4}$ & [-2.20;2.20]$\times 10^{-4}$\\
10\%&[ -8.06; 8.06]$\times 10^{-3}$ &[-4.43;4.43]$\times 10^{-2}$& [-8.84;8.84]$\times 10^{-4}$ \\
\end{tabular}
\end{ruledtabular}
\end{table}

\section{Conclusions}

The Future Circular Hadron Collider with its high center of mass energy and luminosity, as well as novel developments in detector technologies, comes in front of the search for clues to explain the physics beyond SM. Tri-photon production in the hadron colliders provides an ideal platform to search for deviations from SM since it is rare in the SM and involves only pure electroweak interaction contributions at tree level.

In this paper, we have studied anomalous quartic gauge couplings via tri-photon production at FCC-hh. FCC-hh is assumed to operate with a proton beam energy of 50 TeV and an integrated luminosity of 30 ab$^{-1}$. Among the aQGCs where the dimension-8 EFT operators contribute in the $\gamma\gamma\gamma Z$ and $\gamma\gamma\gamma\gamma$ quartic vertices, we have focused on aQGCs of $\mathcal{O}_{T,8}$ and $\mathcal{O}_{T,j}$ opertors more sensitive to tri-photon production. Therefore, cross section is calculated as a function of pure anomalous aQGCs contribution and the interference between the SM process and anomalous aQGCs in order to show the sensitivity of aQGCs $f_{T8}/\Lambda^{4}$ and $f_{T9}/\Lambda^{4}$ to the $pp\to \gamma\gamma\gamma$ production. Since the signal final state has at least three photons, the background processes with the same final state as the signal as well as three dominant background processes relevant to our signal are considered and simulated at the detector level. We also investigate the violation arising from the unitarity in the presence of the dimension-8 aQGCs of our interest.
To investigate the potential enhancement of sensitivity for $f_{T8}/\Lambda^{4}$ and $f_{T9}/\Lambda^{4}$, we conducted a multivariate analysis that incorporated 17 kinematic variables such as transverse momentum and pseudo-rapidity of three-pT ordered photons ($\gamma_1$,$\gamma_2$ and $\gamma_3$), the difference in the azimuthal angle and in pseudo-rapidity between pairs of photons, the distance between pairs of photons in $\eta-\phi$ plane and the invariant mass of three photon system ($m_{\gamma_1\gamma_2\gamma_3}$ ) as input to the TMVA package to perform multivariate analyses via the Boosted Decision Trees (BDT) algorithm. Finally, the threshold of the clipping energy value for each of the aQGC parameters is applied to the invariant mass distributions of the three photon systems.

We have reported limits for $3\sigma$, $5\sigma$ and 95\% C.L., as well as without and with $\delta \to 10\%$ systematic
uncertainty on dimension-eight anomalous quartic gauge couplings $f_{T8}/\Lambda^{4}$ and $f_{T9}/\Lambda^{4}$ with an
integrated luminosity L$_{int}$=30 ab$^{-1} $ and a 100 TeV center of mass energy. Our obtained 95\% C.L. limits on $f_{T8}/\Lambda^{4}$ and $f_{T9}/\Lambda^{4}$ couplings without systematic uncertainties by applying the clipping technique are three orders of magnitude stronger than the best limits recently reported by ATLAS, obtained with an integrated luminosity of 139 fb$^{-1}$ at $\sqrt=13$ TeV. Considering 10\% systematic error, our obtained results at 5$\sigma$ and 95\% C.L. are still two orders of magnitude better than those recently reported by ATLAS.

\end{document}